\documentclass{Maimone}
\usepackage{graphicx}
\usepackage{hyperref}
\usepackage[]{natbib}  
\usepackage{epstopdf}

\def\BibTeX{{\rm B\kern-.05em{\sc i\kern-.025em b}\kern-.08em
    T\kern-.1667em\lower.7ex\hbox{E}\kern-.125emX}}
\bibpunct{(}{)}{;}{a}{}{,}  

\usepackage{aas_macros} 

\begin{document}

\TitreGlobal{SF2A 2022}


\title{The Study of Atmosphere of Hot Jupiters and Their Host Stars}


\author{M.~C. Maimone}\address{Universit\'e C\^ote d'Azur, Observatoire de la C\^ote d'Azur, CNRS, Lagrange, CS 34229, Nice, France}


\author{A. Chiavassa$^1$}

\author{J. Leconte}\address{Laboratoire d'astrophysique de Bordeaux, Univ. Bordeaux, CNRS, B18N, all\'ee Geoffroy Saint-Hilaire, 33615 Pessac, France}



\setcounter{page}{237}


\maketitle


\begin{abstract}
What makes the study of exoplanetary atmospheres so hard is the extraction of its tiny signal from observations, usually dominated by telluric absorption, stellar spectrum and instrumental noise. The High Resolution Spectroscopy has emerged as one of the leading techniques for detecting atomic and molecular species (\citealt{2018arXiv180604617B}), but although it is particularly robust against contaminant absorption in the Earth’s atmosphere, the non-stationary stellar spectrum — in the form of either Doppler shift or distortion of the line profile during planetary transits — creates a non-negligible source of noise that can alter or even prevent the detection. Recently, significant improvements have been achieved by using 3D, radiative hydrodynamical (RHD) simulations for the star and Global Circulation Models (GCM) for the planet (e.g., \citealt{2019A&A...631A.100C}, \citealt{2019AJ....157..209F}). However, these numerical simulations have been computed independently so far, while acquired spectra are the result of the natural coupling at each phase along the planet orbit. With our work, we aim at generating emission spectra of G,F, and K-type stars and Hot Jupiters and coupling them at any phase of the orbit. This approach is expected to be particularly advantageous for those molecules that are present in both the atmospheres (e.g., CO) and form in the same region of the spectrum, resulting in mixed and overlapped spectral lines. We also present the analysis of transmission spectra of the Hot Saturn WASP-20b, observed in the K-band of the recently upgraded spectrograph VLT/CRIRES+ at a resolution R $\sim$ 92, 000 during the first night of the Science Verification of the instrument and that led to a tentative detection of H$_2$O.

\end{abstract}


\section{Introduction}
\begin{figure}[ht!]
 \centering
 \includegraphics[width=0.3\textwidth,clip]{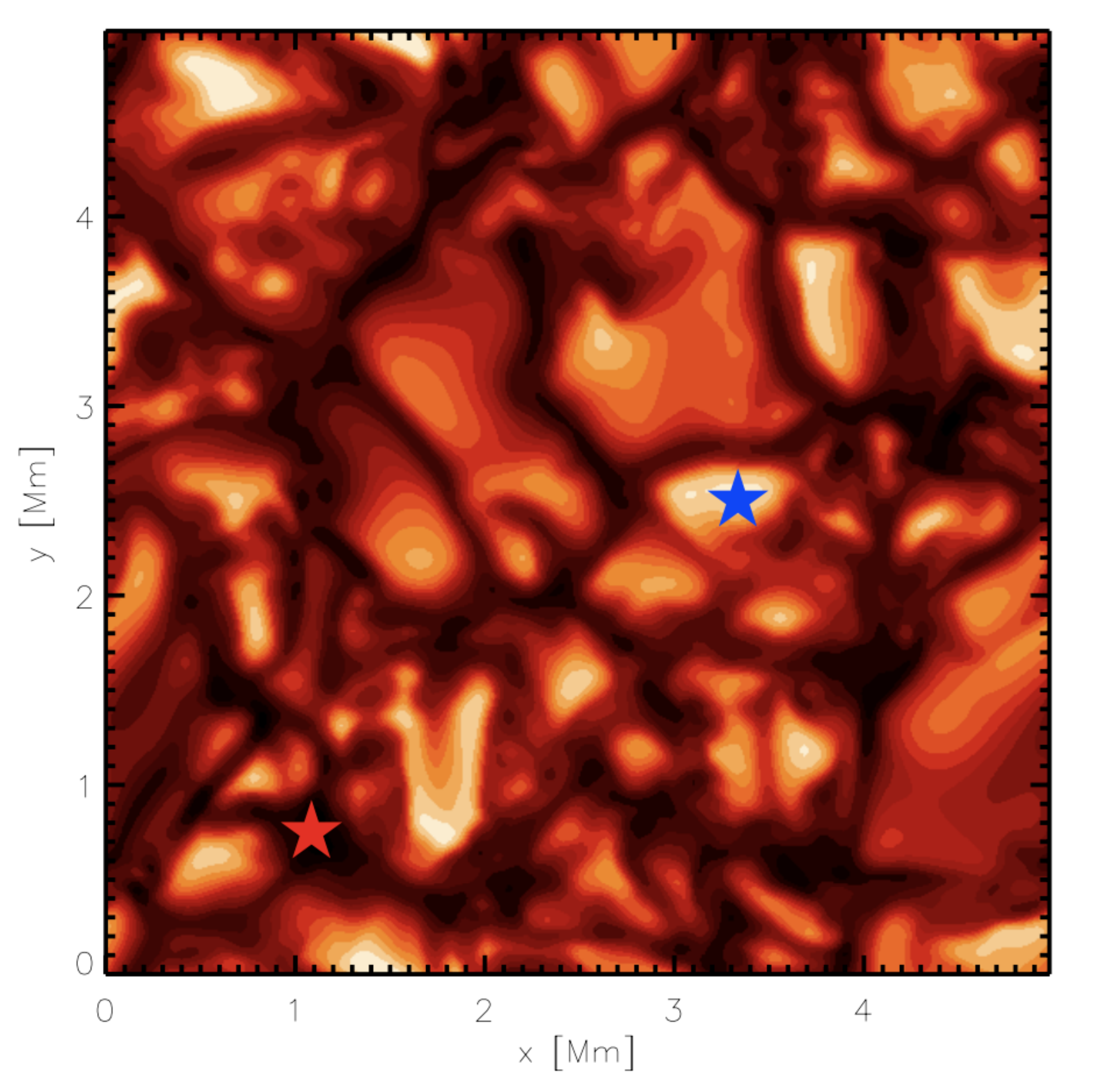}
 \includegraphics[width=0.36\textwidth,clip]{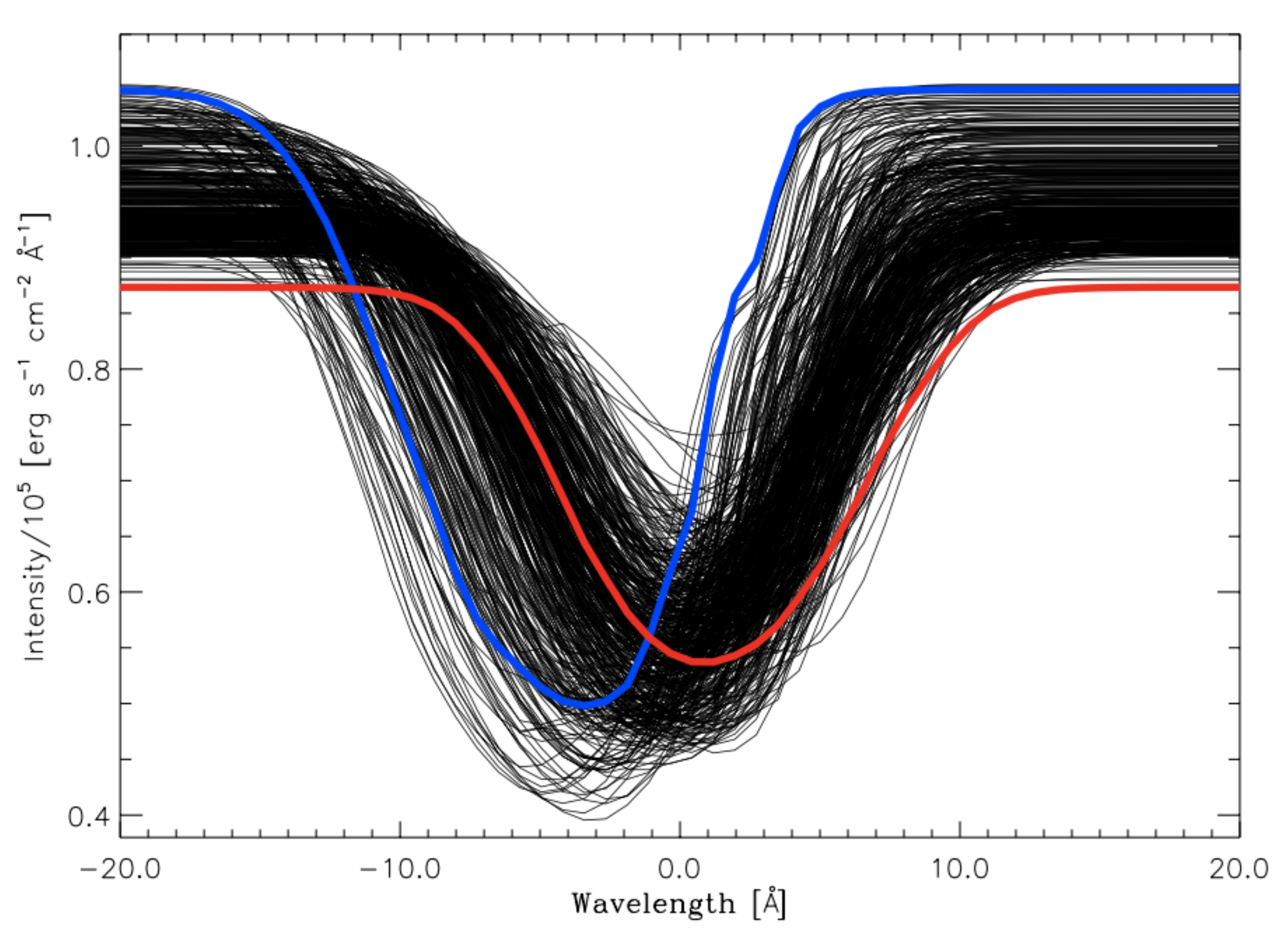}\\
 \includegraphics[width=0.5\textwidth,clip]{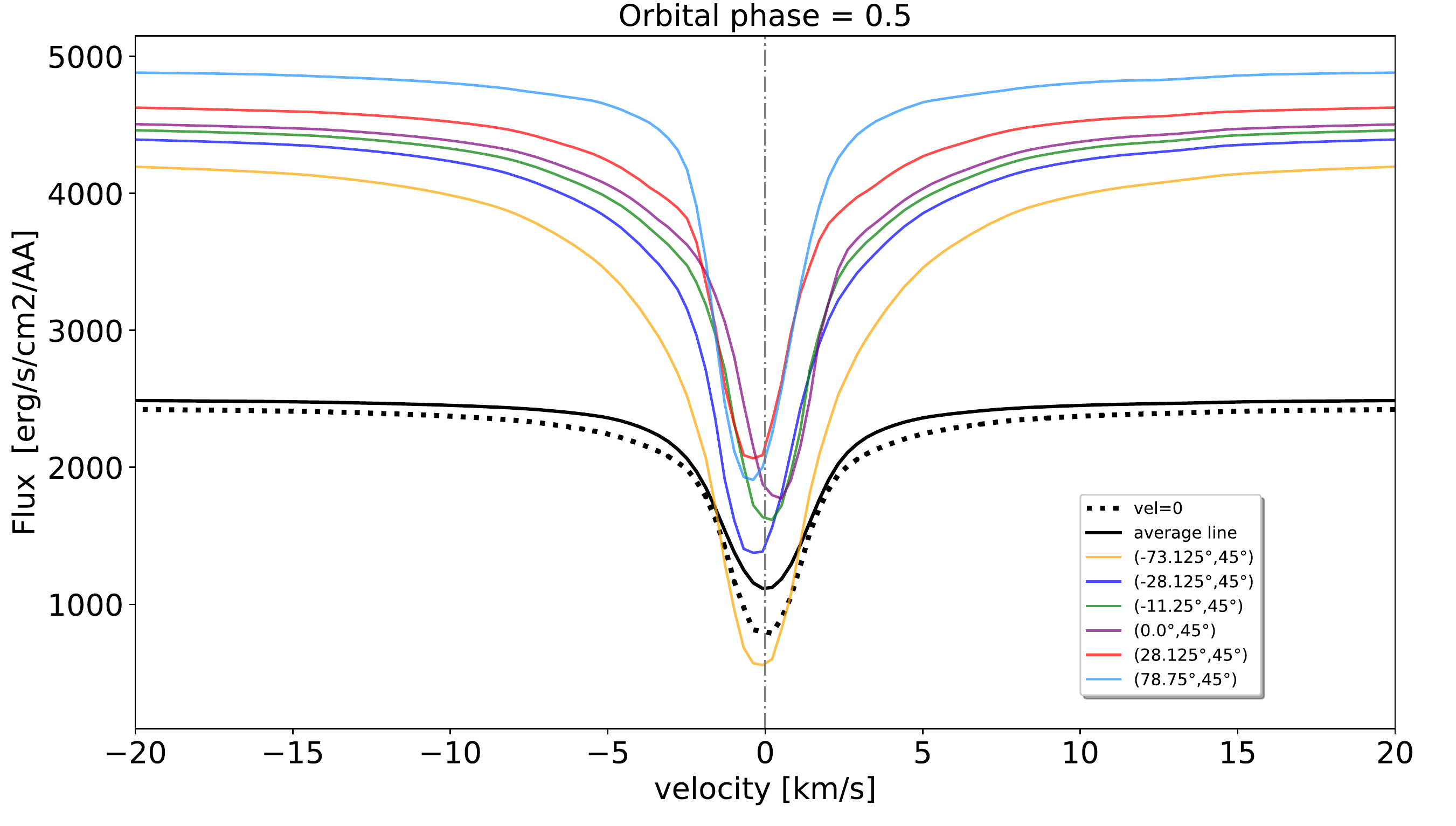}    
  \includegraphics[width=0.27\textwidth,clip]{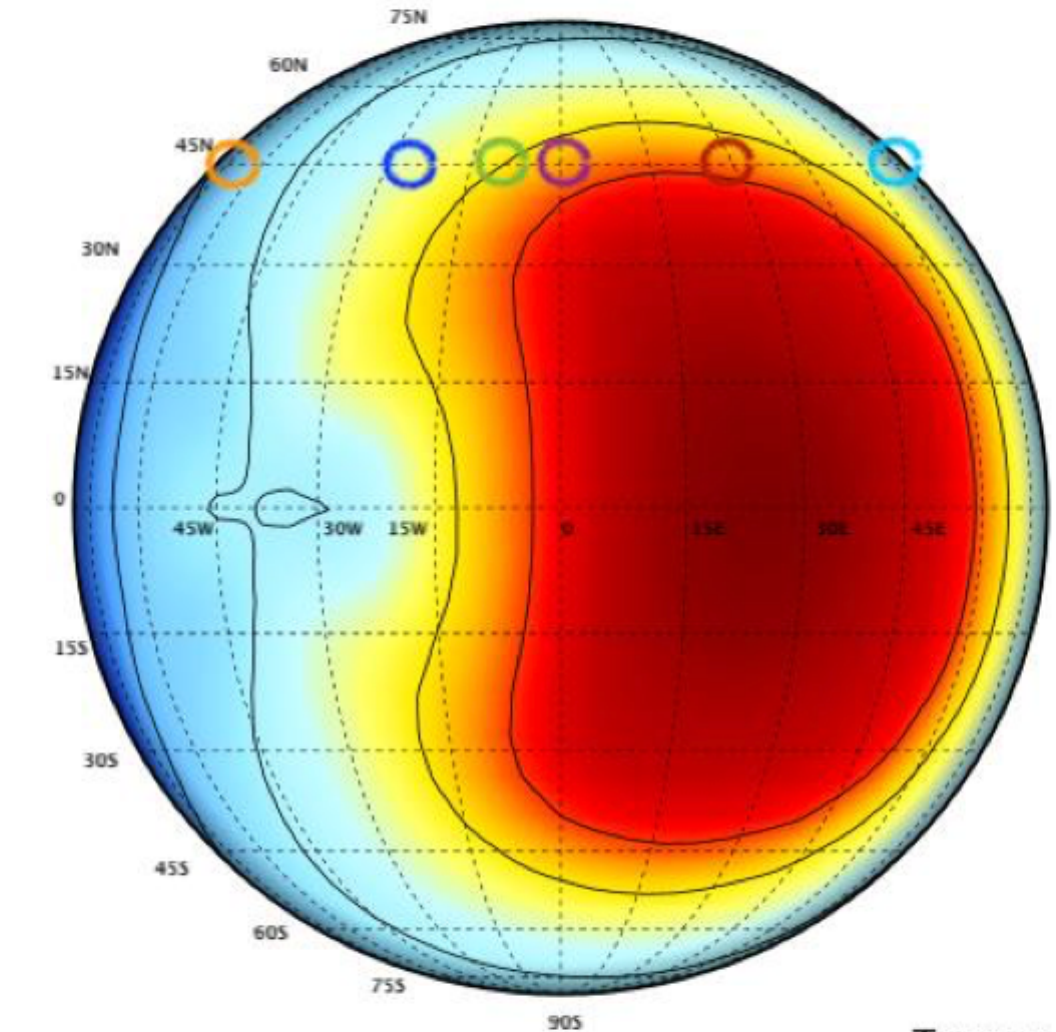}
  \caption{\emph{Top panels: } Intensity maps of a K-dwarf simulation at disk center (left) and the corresponding spatially resolved profiles of one CO line across the granulation pattern (\citealt{2019A&A...631A.100C}) (right). The solid red (intergranular lane) and blue (granule) lines displays two particular positions (colored star symbols) extracted from the intensity map. \emph{Bottom panels: } One particular Hot Jupiter CO line computed for different atmospheric regions (left), which are marked by circles with same colors in the corresponding intensity map (right). }
  \label{maimone:fig1}
\end{figure}

High-resolution spectroscopy (HRS) at resolving powers $R>50\,000$ has proved to be one of the leading technique for remote atmospheric characterisation of exoplanet atmospheres.
HRS allows to partially resolve the molecular dense forest of lines and to robustly identify them through line-matching techniques such as cross-correlation (\citealt{2010Natur.465.1049S}).
This technique occurred to be particularly suitable in the Near-Infrared due to a favourable planet-to-star flux contrast ratio and to the presence of strong absorption bands of the main carries of carbon (CO) and oxygen (H2O) (\citealt{2012EGUGA..1413720M}).\\
Although several molecules (CO, H$_2$O, CH$_4$, HCN, TiO) have been detected in the atmosphere of a dozen exoplanets (\citealt{2018arXiv180604617B} for a review), characterizing an exoplanet remains a challenge. 
Stars are non-uniform and can be potential source of spurious signals, which can alter or even prevent the interpretation of exoplanet spectra \citep{2017A&A...597A..94C,2019ApJ...879...55C,2019A&A...631A.100C,2021A&A...649A..17D}. In particular, the granulation pattern associated with the heat transport by convection \citep{2009LRSP....6....2N} has a temporal and spatial variability which manifests as Doppler shifts and distortions of the line profile (Fig.~\ref{maimone:fig1}, top panels) during the planetary transit. Similarly,  atmospheric circulation and planetary winds affect the planetary spectrum (Fig.~\ref{maimone:fig1}, bottom panels), causing an overlap of effects of similar time-scales  (\citealt{2019AJ....157..209F}). \\ 
In this context, the use of 3D, radiative hydrodynamical (RHD) simulations for stars  and Global Circulation Models (GCM) for planets play a major role (e.g., \citealt{2019A&A...631A.100C} and yet, so far they have been computed independently, while acquired spectra are the result of the natural coupling at each phase along the planet orbit.

\section{The two side of the project}

\subsection{Emission spectroscopy: synthetic observables and the study of detectability of the planet signal}

\begin{figure}[ht!]
 \centering
 \includegraphics[width=0.28\textwidth,clip]{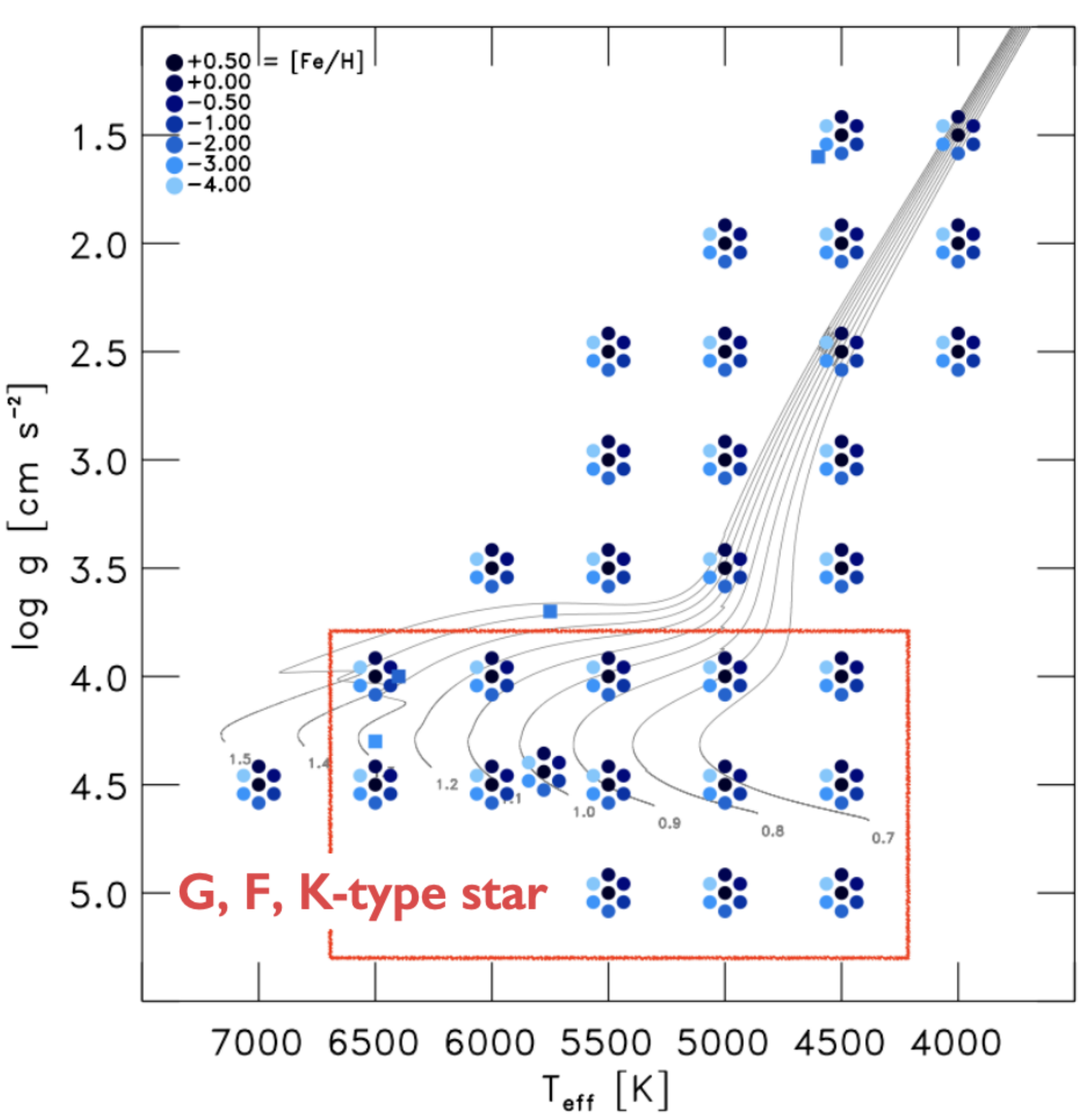}
 \includegraphics[width=0.35\textwidth,clip]{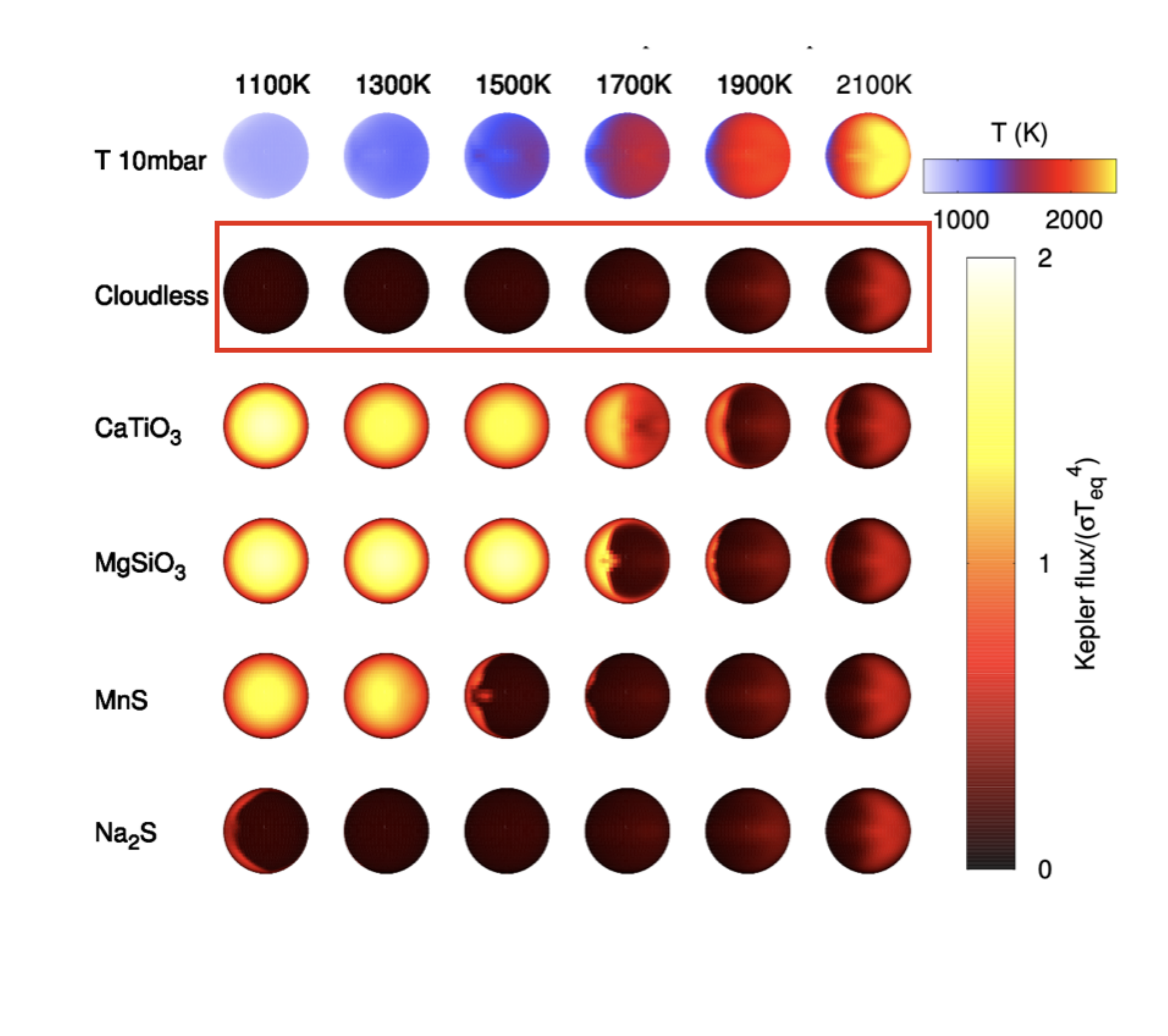}  
  \includegraphics[width=0.35\textwidth,clip]{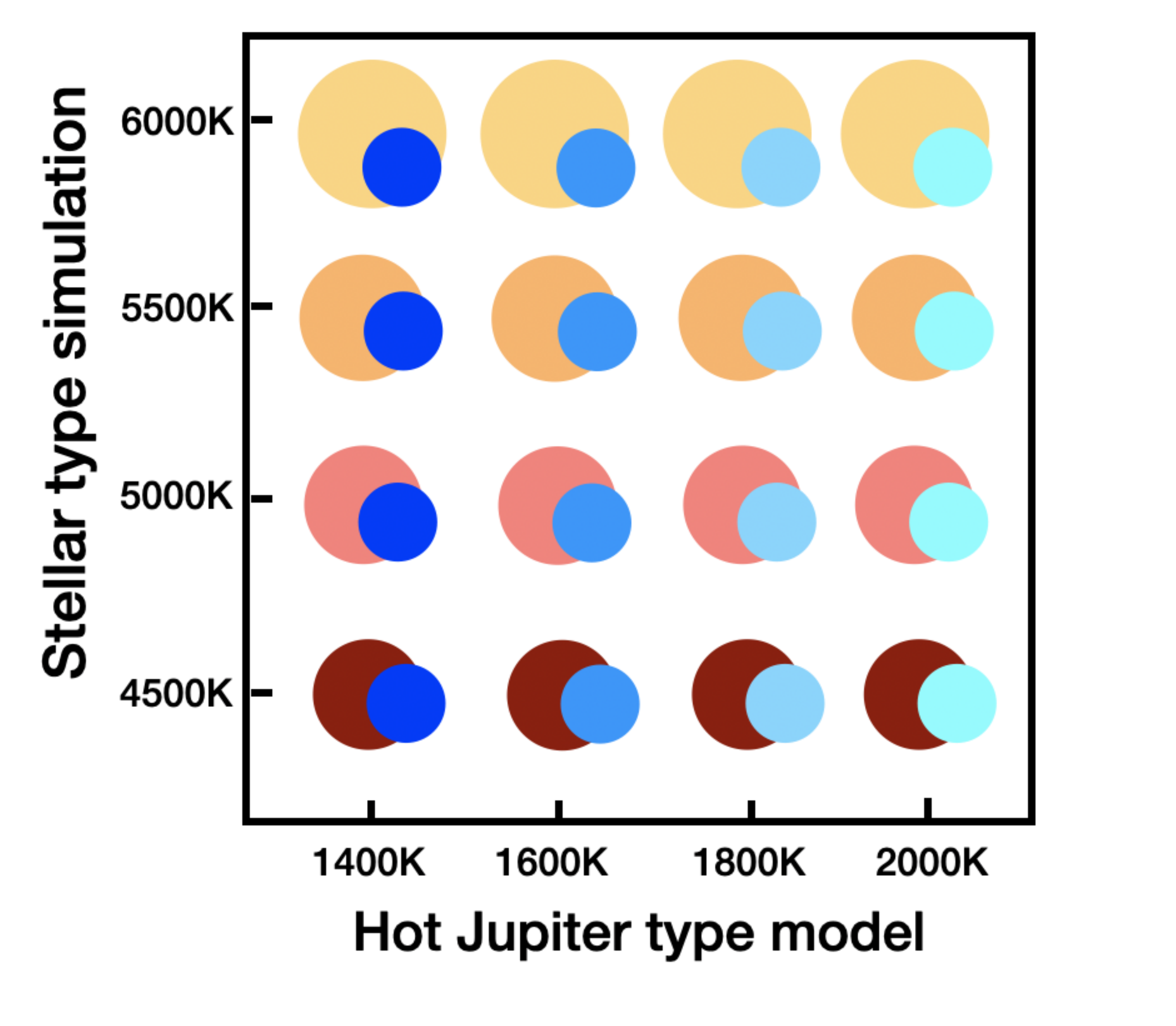}  
  \caption{\textit{Left: }3D RHD simulation-grid in the H-R diagram computed with \textsc{Stagger-code} (\citealt{2013A&A...557A..26M}. In particular, the G,F and K stellar types (within the red line) are the ones used in this work. \textit{Central: } Temperature and outgoing flux from the dayside of Hot Jupiters with different equilibrium temperatures and cloud species (\citealt{2016ApJ...828...22P}). Cloud-free models (within the red line) are the ones used in this work. \textit{Right: } Estimated coupling grid of those stellar and planet models marked in the previous panels.
  }
  \label{maimone:fig2}
\end{figure}

We are developing a new approach which aims at  generating high resolution emission spectra of Hot Jupiters (1400K-2100K) and their host star (G,F,K-type) combined along the orbit to reproduce realistic synthetic observations. The innovation and uniqueness of this approach is the use of 3D hydrodynamical simulations for the atmosphere of both the star (\textsc{Stagger-code}, \citealt{2009LRSP....6....2N, 2013A&A...557A..26M}) and the planet \citep[MITgcm][]{2013ApJ...762...24S,2021MNRAS.501...78P}. Two summary grids are given in Fig.~\ref{maimone:fig2}, left and central panels, respectively. The coupling is done by using an updated version of the post-processing radiative transfer code \textsc{Optim3D} \citep{2009A&A...506.1351C}. The code takes into account, simultaneously, the stellar and planetary dynamics, which influence the shape, shift, and asymmetries of spectral lines, and not rarely of the same specie.  \\ 
We will perform a study of detectability of the planet signal by exploring the coupling grid (right panel of Fig.~\ref{maimone:fig2}) to understand if any other star-planet couple can reproduce the same signal of the input one. \\
With the new generation of telescopes (e.g., CAHA/CARMENES, VLT/CRIRES+, CHFT/SPIRou, ELT, etc,...) and the increasing number of planets expected to be observed, our tool will be extremely useful to interpret HRS data to extract information either for the characterization of the stellar parameters and metallicity and for the planet dynamics and composition.

\subsection{Transmission spectroscopy: tentative detection of H$_2$O in the atmosphere of WASP-20b}

\begin{figure}[hb!]
 \centering
 \includegraphics[width=0.45\textwidth]{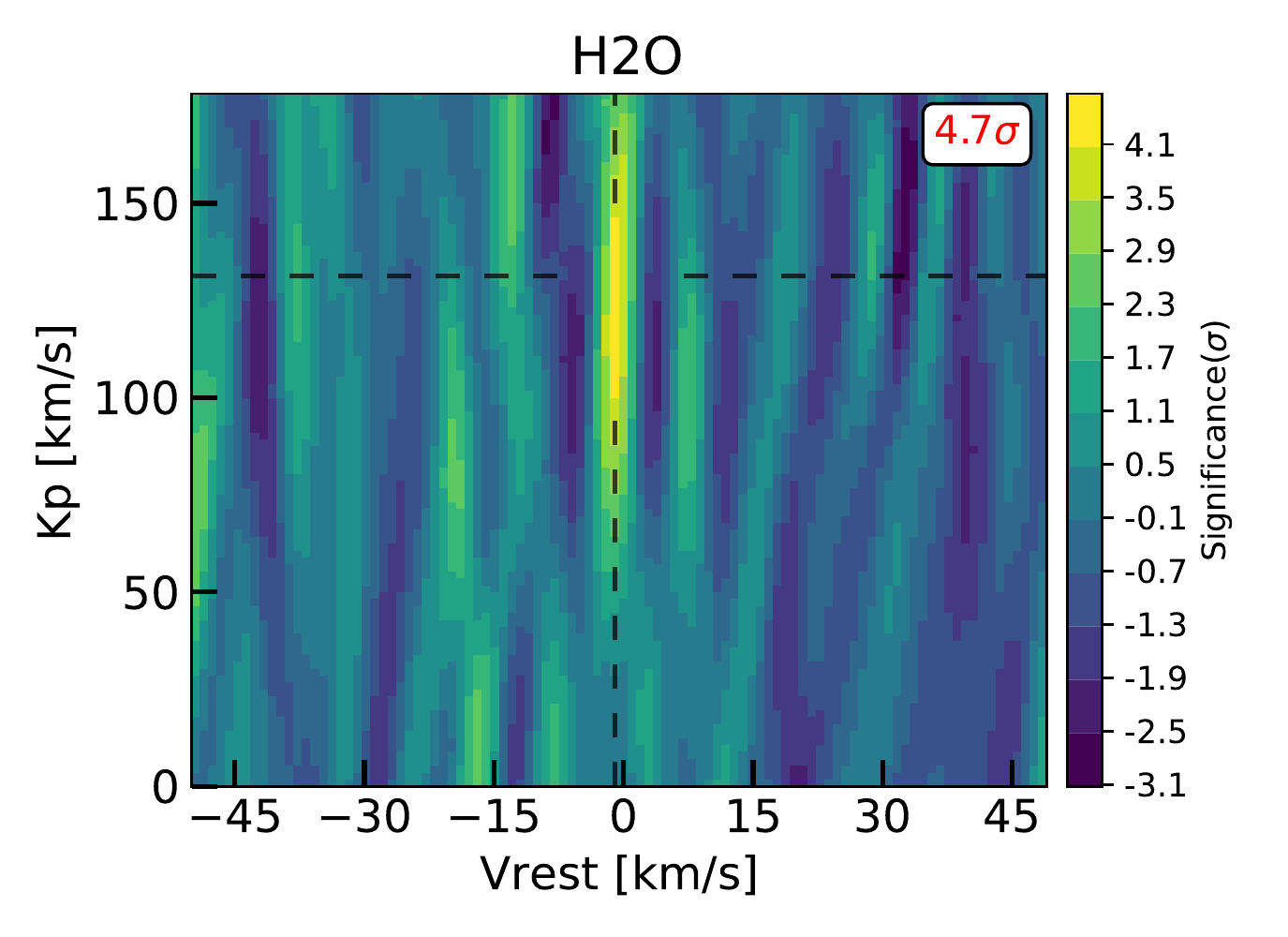} 
 \includegraphics[width=0.45\textwidth]{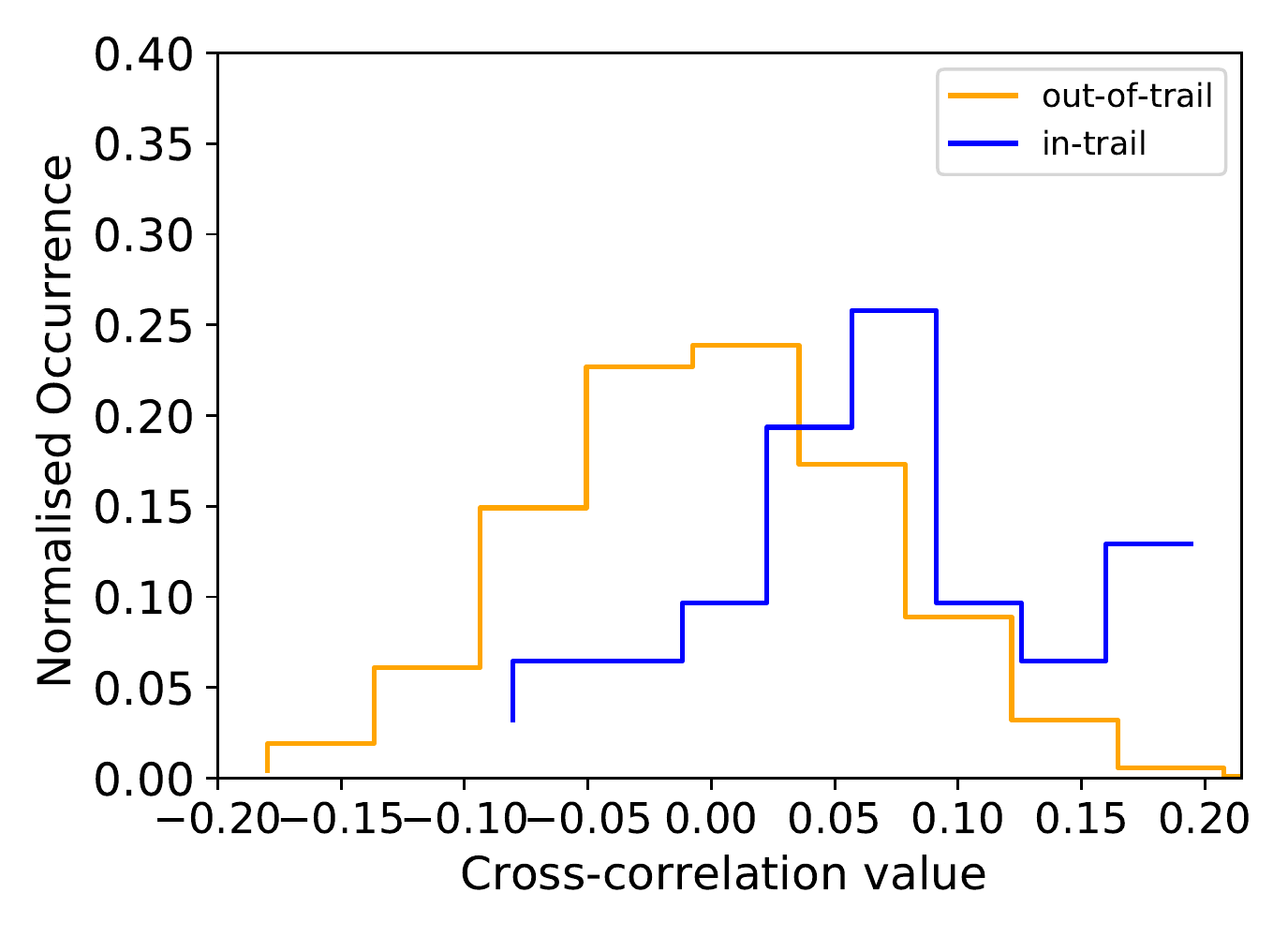} 
  \caption{\textit{Left: }Total cross correlation signal from the GCM used for the atmosphere of WASP-20b in our analysis, shown as a function of rest-frame velocity and planet projected orbital velocity.  The GCM is cloud-free, in chemical equilibrium with no thermal inversion and contains a VMR[H$_2$O] equals to $-3.3$. \textit{Right: } Comparison between out-of-trail (yellow solid line) and in-trail (blue solid line) cross-correlation distributions of WASP-20b. The latter is systematically shifted towards higher.  A Welch t-test on the data rejects the hypothesis that the two distributions are drawn from the same parent distribution at the 4.1$\sigma$.
      }
  \label{maimone:fig3}
\end{figure}

We present the analysis of transmission spectra of the inflated, Saturn-mass planet WASP-20b (0.31 M$\rm{_J}$; 1.46 R$\rm{_J}$) orbiting a F9-type star in less than 5 days and with an equilibrium temperature of $\sim$1400K (Maimone et al. 2022, submitted).  Our observations occurred on Sept, 16 2021, during the very first night of the Science Verification of the recently upgraded spectrograph VLT/CRIRES+ (\citealt{2016SPIE.9908E..0ID}). Our goal was a demonstration of the basic capabilities of the new instrument and we chose to observe WASP-20 because it was the only target with a visible transit during the SV observing window. The system was observed for the first time by \cite{2015A&A...575A..61A}, discovered as a binary system separated by only 0.26" by \cite{2016ApJ...833L..19E} and confirmed by \cite{2020A&A...635A..74S}. \\
Because with CRIRES+ we did not resolve the binary nature of WASP-20 found by \cite{2016ApJ...833L..19E} and \cite{2020A&A...635A..74S}, we treated the system as a single star and took as main reference \cite{2015A&A...575A..61A}.
We used Principal Component Analysis (PCA,\citealt{1987ASSL..131.....M},\citealt{1992nrfa.book.....P}) to remove any dominant time dependent contaminating features (such as telluric bands, stellar absorption lines and systematic instrumental trends) and we cross-correlated the residual spectra with models of the WASP-20b atmosphere to extract the planet spectrum, as done by \cite{2021Natur.592..205G}.
We used synthetic spectra computed from 1D models using GENESIS (\citealt{2017MNRAS.472.2334G}), and from SPARC/MIT Global Circulation Models (GCM, \citealt{2013ApJ...762...24S}, \citealt{2021MNRAS.501...78P}, \citealt{2022A&A...658A..42P}) using Pytmosph3R (\citealt{2019A&A...623A.161C}, \citealt{2022A&A...658A..41F}). The best fitting model was a cloud-free GCM at 1400K containing only H$_2$O (VMR=-3.3), in chemical equilibrium and with no thermal inversion. The cross-correlation matrix (shown in Fig.~\ref{maimone:fig3}, left panel) peaks at K$_{\rm{P}}$=131$^{+23}_{-39}$ km s$^{-1}$ and V$_{\rm{rest}}$=$-1\pm{1}$ km s$^{-1}$, in accordance with the RV semi-amplitude expected from \cite{2015A&A...575A..61A} results. \\
We determined the statistical significance of the H$_2$O signal as in previous works (\citealt{2012Natur.486..502B}, \citealt{2013ApJ...767...27B}). From the matrix containing the cross correlation signal as function of planet radial velocity and time, CCF(V, t), we selected those values not belonging to the planet RV curve (out-of-trail) and those ones belonging to the planetary trace (in-trail), as shown in Fig.~\ref{maimone:fig3}, right panel. A Welch T-test ruled out the in-trail and out-of-trail values having been drawn from the same parent distribution at the 4.1$\sigma$ level. More details will be found in Maimone et al. 2022 (submitted).


\begin{acknowledgements}
This project has received funding from the European Research Council (ERC) under the European Union’s Horizon 2020 research and innovation programme (grant agreement n◦ 679030/WHIPLASH).
\end{acknowledgements}

\bibliographystyle{aa}  
\bibliography{Maimone} 

\end{document}